\documentclass[aip,amsmath,amssymb,reprint]{revtex4-1}

\usepackage{graphicx}%
\usepackage{gensymb}
\usepackage{dcolumn}%
\usepackage{bm}%
\usepackage[utf8]{inputenc}
\usepackage[T1]{fontenc}
\usepackage{etoolbox}
\usepackage{braket}
\usepackage{siunitx}
\usepackage{amsmath}
\usepackage{upgreek}

\begin{document}
\title{Towards Quantum Sensing of Chiral-Induced Spin Selectivity:\\ Probing Donor--Bridge--Acceptor Molecules with NV Centers in Diamond}
\author{Laura A. Völker}
\altaffiliation{Equal contribution}
\author{Konstantin Herb}
\altaffiliation{Equal contribution}
\author{Erika Janitz}
\author{Christian L. Degen}
\author{John M. Abendroth}
\email{jabendroth@phys.ethz.ch}

\affiliation{ 
1 Department of Physics, ETH Zurich, Otto-Stern-Weg 1, 8093 Zurich, Switzerland
}

\date{February 3 2023}

\begin{abstract}
Photoexcitable donor–bridge–acceptor (D--B--A) molecules that support intramolecular charge transfer are ideal platforms to probe the influence of chiral-induced spin selectivity (CISS) in electron transfer and resulting radical pairs. In particular, the extent to which CISS influences spin polarization or spin coherence in the initial state of spin-correlated radical pairs following charge transfer through a chiral bridge remains an open question. Here, we introduce a quantum sensing scheme to measure directly the hypothesized spin polarization in radical pairs using shallow nitrogen--vacancy (NV$^-$) centers in diamond at the single- to few-molecule level. Importantly, we highlight the perturbative nature of the electron spin–spin dipolar coupling within the radical pair, and demonstrate how Lee-Goldburg decoupling can preserve spin polarization in D--B--A molecules for enantioselective detection by a single NV$^-$ center. The proposed measurements will provide fresh insight into spin selectivity in electron transfer reactions. %
\end{abstract}
\hspace{0.5cm}
\maketitle

Chiral-induced spin selectivity (CISS) describes electron-spin-dependent and enantioselective interactions in charge transport through chiral molecules.\cite{Evers2022,Aiello2022} Recently, donor--bridge--acceptor (D--B--A) molecules that support intramolecular charge transfer resulting in spin-correlated radical pairs (SCRPs) have emerged as ideal systems to probe the role of CISS in photoinduced electron transfer.\cite{Carmeli2014, Abendroth2019, Junge2020, Fay2021, Fay2021b, Luo2021, Chiesa2021, Privitera2022} As radical pairs are relevant both for biology\cite{Grissom1995, Hore2016} and for quantum information science using spin qubit pairs,\cite{Harvey2021, Mani2022} unravelling the influence of bridge chirality on the initial spin state is of particular importance. Proposals to test CISS in electron transfer reactions within D--B--A molecules have included the use of electron paramagnetic resonance (EPR) spectroscopy on the radical pair itself \cite{Fay2021, Luo2021} as well as EPR and nuclear magnetic resonance (NMR) spectroscopy of a strongly coupled reporter spin. \cite{Chiesa2021} However, conventional EPR or NMR methods described by these strategies are challenged by low sensitivities that preclude studies on dilute molecular assemblies. Furthermore, proposed experiments necessitate fully oriented chiral molecular assemblies, which poses experimental difficulties.

Alternatively, quantum sensing with the negatively charged nitrogen--vacancy (NV$^-$) center in diamond provides a route to detect minute magnetic fields with high sensitivity and nanoscale spatial resolution.\cite{Meirzada2021,Janitz2022} When positioned close ($<10$ {nm}) to the diamond surface, these fluorescent defects can be leveraged for chemical sensing of magnetic fields in proximal molecules.\cite{Staudacher2013,Mamin2013,Liu2022, Xie2022} Moreover, the diamond surface provides a natural platform to bind and orient anchored molecules, facilitating NV$^-$ measurements in the single- to few-molecule regime.\cite{Sushkov2014,Lovchinsky2016,Abendroth2022} Notably, the use of NV$^-$ centers for investigating radical pairs has been explored theoretically,\cite{Liu2017,Finkler2021} highlighting the exciting potential to monitor their spin dynamics in dilute systems.

Here, we propose optically detected magnetic resonance (ODMR) spectroscopy with single NV$^-$ centers in diamond to measure directly the hypothesized spin polarization attributed to CISS in photoexcited D--B--A molecules (\textbf{Fig. 1}). We first introduce a general framework to describe the initial spin state of a radical pair born from CISS. Focusing on spin-polarized states, we expose the perturbing nature of the pseudosecular spin--spin dipolar coupling of the radical pair, and leverage Lee-Goldburg decoupling as one possible solution to preserve spin polarization for readout by ODMR. We establish critical measurement design rules to overcome geometrical limitations of dipolar sensing with the NV$^-$, and evaluate the detection sensitivity for spin-polarization resulting from CISS in experimentally relevant systems. Finally, we provide perspective on how quantum sensing approaches are optimal to unambiguously describe  the spin state of an electron following transport across a chiral bridge.

\begin{figure*}
\includegraphics[width=1\textwidth]{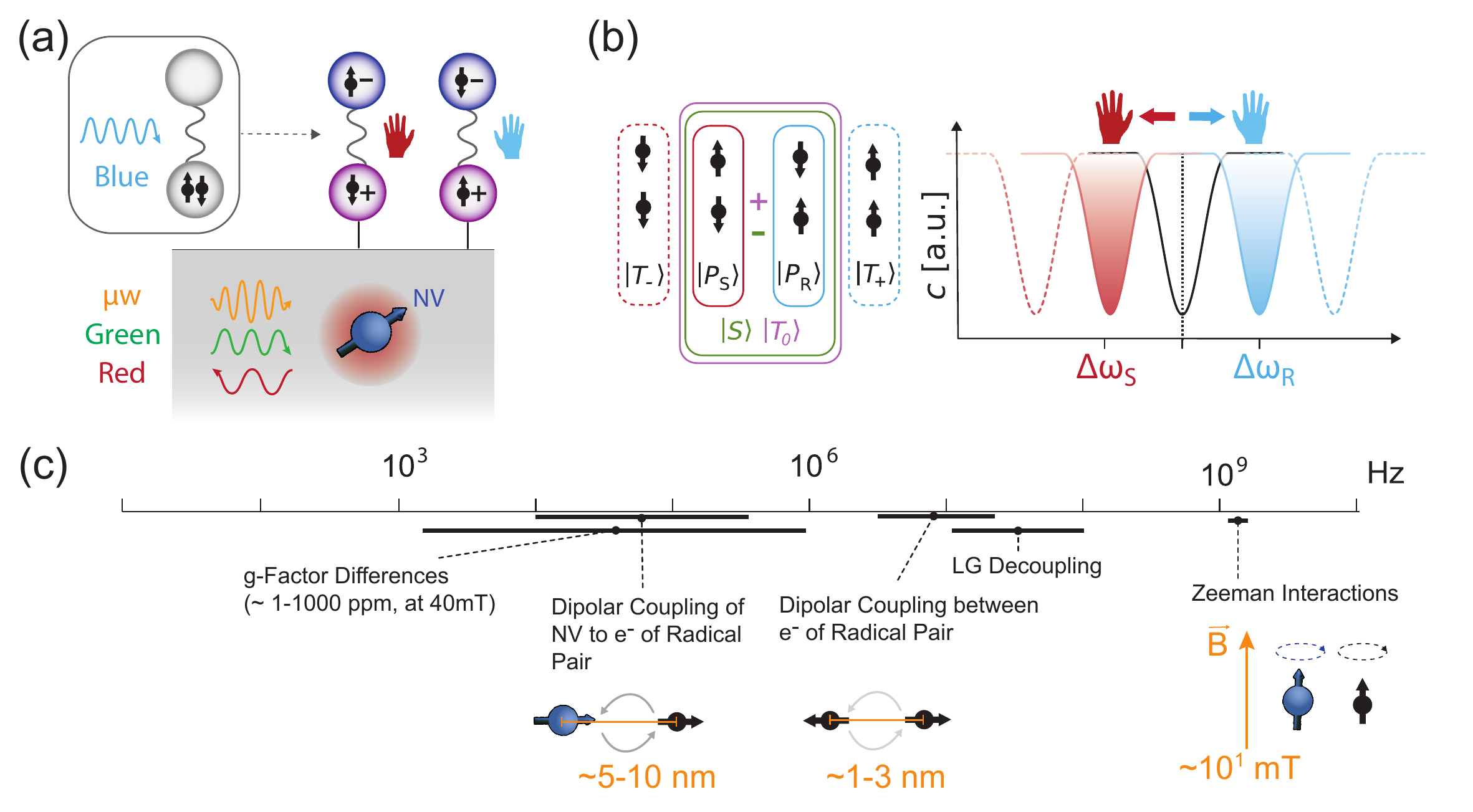}
\caption{\label{fig:1} \textbf{(a)} Schematic of the proposed experiment: Donor--bridge--acceptor (D--B--A) molecules are anchored onto the surface of a diamond containing shallow NV$^-$ centers. Radical pairs are excited upon irradiation with a blue laser inducing charge transfer through an achiral or chiral bridge. The spin state of the radical pairs is probed with the NV$^-$ by means of optically detected magnetic resonance (ODMR), which includes optical spin-state preparation and readout of the quantum sensor as well as coherent spin-state manipulations with microwave ($\upmu$w) irradiation. \textbf{(b)} Schematic representation of the Zeeman basis states (left) and hypothesized result of the experiment (right). The dipolar coupling to the radical pair induces a frequency shift on the NV$^-$ center's resonance, with a magnitude that depends on the spin state of the radical pair, \textit{i.e.}, relative occupations of the Zeeman basis states. Spin polarization attributed to CISS would reveal itself as an asymmetric frequency shift that changes sign upon reversal of the bridge chirality in D--B--A molecules. \textbf{(c)} Relative orders of magnitude of interactions in a three-spin system consisting of the electron spin of a NV$^-$ center and a single radical pair. Dipolar couplings are provided for a range of expectable distances and all Zeeman effects are evaluated at relevant magnetic fields for this sensing protocol. At magnetic field strengths of several tens of mili-Tesla, the Zeeman interaction is the dominant interaction such that the high-field approximation holds.}
\end{figure*}

Charge transfer and radical pair formation in photoexcited D--B--A molecules is usually a spin-conserving process. Indeed, for an achiral bridge, the charge separation preserves the molecular singlet state $\ket{S} = \frac{1}{\sqrt{2}} \left[\ket{\uparrow  \downarrow} -\ket{\uparrow  \downarrow}\right]$. Interconversion of this singlet with the triplet $\ket{T_0} = \frac{1}{\sqrt{2}} \left[\ket{\uparrow  \downarrow} + \ket{\uparrow  \downarrow}\right]$ state (commonly termed "zero-quantum coherence") is driven by asymmetries in the local spin environment of the two electrons in the charge-separated state. This intersystem crossing has been at the heart of radical pair research for decades, and is crucial for spin-correlated\footnote{For spin-correlated radical pairs, the measurement of one electron spin fully determines the measurement result of the second electron spin. However, correlations do not enforce that each repetition yields the same measurement result, \textit{e.g.}, for a singlet state. Alternatively, for a spin-polarized state, as potentially induced by CISS, the measurement should yield (in the appropriate basis) always the same result, and a therefore also a correlation.} and conventional radical-pair mechanisms where coherence is preserved or lost, respectively.\cite{Steiner1989,Zarea2015} However, if the bridge is chiral, whether or not CISS plays a role is an outstanding question. Namely, can CISS influence spin polarization or spin coherence in the initial state?

We work under the assumption of a spin-conserving electron transfer and in a coordinate system aligned with the main quantization axis defined by the molecular axis $\vec{a}_{\mathrm{RP}}$. We start with the initial spin state of the radical pair $\ket{\psi_0}$ that resides in the zero-quantum subspace \footnote{Notably, the boundary condition of a spin conserving electron transfer does not necessarily exclude cases where elements in the double-quantum subspace of $\rho_0$ are non-zero.  However, it dictates that this population must occur in a symmetric fashion, \textit{i.e.}, $\bra{T_+}\rho_0\ket{T_+} = \bra{T_-}\rho_0\ket{T_-}$. As long as these states are coherent, such double-quantum subspace populations can be  reduced to the zero-quantum subspace, allowing a description of the state using Eq. \eqref{eq:initial_state}, by means of a coordinate transformation. Any coordinate transformation on  Eq. \eqref{eq:initial_state} does not affect the $\ket{S}$ contributions of the state, but transforms the triplet states into one another such that $\ket{T_0}$ is depopulated and $\ket{T_+}, \ket{T_-}$ are populated in a symmetric fashion. Essentially, it mimicks the evolution of the state under a microwave pulse with angles and phases corresponding to the azimuthal and polar angles of the coordinate system transformation. During the transformation, the polarization $p$ of the state is reduced by $\cos{\theta}$, with $\theta$ being the azimuthal angle of the transformation.} and is formulated in the singlet--triplet basis in accordance with recent theoretical work \cite{Fay2021, Fay2021b, Luo2021, Chiesa2021} 
\begin{equation}
\ket{\psi_0} = \cos{\alpha}\ket{S} + e^{i \beta}\sin{\alpha}\ket{T_0}.
\label{eq:initial_state}
\end{equation}
The parametrization in $\alpha$ and $\beta$ allows for all possible coherent superpositions of $\ket{S}$ and $\ket{T_0}$. However, the influence of CISS on the initial state can be described more intuitively in the Zeeman basis $\left\{ \ket{T_+} = \ket{\uparrow\uparrow}, \ket{P_\mathrm{R}} = \ket{\uparrow\downarrow}, \ket{P_\mathrm{S}}=\ket{\downarrow\uparrow},\ket{T_-}=\ket{\downarrow\downarrow}\right\}$ (\textbf{Fig. \ref{fig:1}b}). Here, $\ket{T_+}$ and $\ket{T_-}$ span the double-quantum subspace (coincident in both bases), and $\ket{P_\mathrm{S}}$ and $\ket{P_\mathrm{R}}$ represent oppositely polarized states that may be populated differently for bridges of opposite handedness \textit{S} and \textit{R}. We formulate the density matrix $\rho_0(\alpha , \beta , \Lambda)=\ket{\psi_0}\bra{\psi_0}$ as 
\begin{equation}
\rho_0 = 
 \begin{pmatrix} 0 & 0 & 0 &  0 \\
 0 & \frac{1+ \sin{2\alpha}{\cos{\beta}}}{2} & \Lambda \frac{\cos{2\alpha}-i \sin{2\alpha}\sin{\beta}}{2} & 0 \\ 
0 & \Lambda \frac{\cos{2\alpha} + i \sin{2\alpha}\sin{\beta}}{2} & \frac{1- \sin{2\alpha}{\cos{\beta}}}{2} & 0 \\
0 & 0 & 0 & 0 
 \end{pmatrix},
\label{eq:initial_rho}
\end{equation} 
where we introduced a damping parameter $\Lambda$ on all off-diagonal elements to account for incoherent states. Using Eq. \eqref{eq:initial_rho}, we can describe all possible initial states using a minimal set of parameters:  
\begin{enumerate}
\item[\textbf{(i)}]  $\alpha \in [-\frac{\pi}{2},\frac{\pi}{2}]$ describes the degree of mixing between $\ket{S}$ and $\ket{T_0}$. We can recover $\ket{S}$ and $\ket{T_0}$ for $\alpha = 0$ and $\alpha =\pm\frac{\pi}{2}$, respectively, while $\ket{P_\mathrm{S}}$ and $\ket{P_\mathrm{R}}$ states are obtained for $\alpha = \pm \frac{\pi}{4}$ when $\beta = 0$.
\item[\textbf{(ii)}] $\beta \in [-\pi, \pi]$ describes the relative phase offset between $\ket{S}$ and $\ket{T_0}$. If $\beta = \pm \frac{\pi}{2}$, $\ket{P_\mathrm{S}}$ and  $\ket{P_\mathrm{R}}$ states are equally populated for all $\alpha$, and the initial state would not be polarized.\cite{Fay2021}
\item[\textbf{(iii)}] $\Lambda \in [0, 1]$ describes the extent of coherence in the state. If $\Lambda = 1$ we obtain a pure, fully coherent state, while if $\Lambda < 1$, a mixed state with reduced coherence (\textit{i.e.,} $Tr(\rho_0^2) < 1$) is present. When $\Lambda = 0$, the state is fully incoherent and resembles common representations of polarized states in NMR and EPR spectroscopy.\cite{Chiesa2021} 
\end{enumerate}

The CISS-induced polarization $p_{\mathrm{CISS}}$, defined as an imbalance in occupations of $\ket{P_\mathrm{S}}$ and $\ket{P_{\mathrm{R}}}$ states, follows as \begin{equation}
p_{\mathrm{CISS}} = \bra{P_{\mathrm{S}}} \rho_0 \ket{P_{\mathrm{S}}} - \bra{P_{\mathrm{R}}} \rho_0 \ket{P_{\mathrm{R}}} = \sin{2\alpha}\cos{\beta}.
\label{eq:polarization}
\end{equation} Thus, $p_{\mathrm{CISS}}$ vanishes for $\ket{S}$ and $\ket{T_0}$ states, and changes sign for $\ket{P_\mathrm{S}}$ and $\ket{P_\mathrm{R}}$ states. The presence of an external magnetic field $\vec{B}=(0, 0, B_z)$, common for sensing with NV$^-$ centers, defines the laboratory frame ($\vec{a}_{\mathrm{B}}$). We do not assume \textit{a priori} that $\vec{a}_{\mathrm{B}}$ is necessarily aligned with $\vec{a}_{\mathrm{RP}}$, such that the apparent polarization $p$ in the laboratory frame is reduced to 
\begin{equation}
p = p_{\mathrm{CISS}}\cos{\theta_{\mathrm{RP}}},
\end{equation}
which depends on the relative orientation $\theta_{\mathrm{RP}}$ between the two frames. In this work, \textit{p} is the relevant figure of merit for the proposed sensing scheme.

Next, we explore the evolution of these states on the relevant timescale of our sensing experiment (\textit{i.e.}, assuming charge-separated state lifetimes on the order of a few $\upmu$s or longer and excluding environmentally driven relaxation effects). Here, we include the evolution of the initially prepared spin state under the Larmor precession of the radical spins that have potentially different \textit{g}-factors $g_1$ and $g_2$ and under the dipolar coupling between them. The former evolution is described by the Zeeman Hamiltonian 
\begin{equation}
\hat{H}_Z^{\mathrm{RP}} = B_z(\gamma_1 \hat{S}_{z}^{(1)} + \gamma_2\hat{S}_{z}^{(2)})
\label{eq:zeeman}
\end{equation}
with gyromagnetic ratios $\gamma_1= \frac{\mu_B}{\hbar} g_1$ and $\gamma_2= \frac{\mu_B}{\hbar}g_2$ where $\mu_B$ is the Bohr magneton and $\hbar$ is the reduced Planck's constant ($\hat{H}$ in angular frequency units, $\hat{H} =\hat{\mathcal{H}}/\hbar$). The latter evolution is described by the dipolar coupling Hamiltonian 
\begin{widetext}
\hspace{0.25cm}
\begin{equation} \begin{split}
\hat{H}_{\mathrm{dip}}^{\mathrm{RP}} &= \frac{\mu_0\hbar \gamma_1 \gamma_2}{4 \pi}\frac{1}{s^3} \left(\vec{\hat{S}}^{(1)}\cdot \vec{\hat{S}}^{(2)} -  \frac{3}{s^2}(\vec{\hat{S}}^{(1)}\cdot \vec{s})(\vec{\hat{S}}^{(2)}\cdot \vec{s})\right)   \\
&= \frac{\mu_0\hbar \gamma_1 \gamma_2}{4 \pi}\frac{(1-3\cos^2{\theta_{\mathrm{RP}}})}{s^3}\left[\hat{S}_z^{(1)}\hat{S}_z^{(2)} + (\hat{S}_+^{(1)}\hat{S}_-^{(2)} + \hat{S}_-^{(1)}\hat{S}_+^{(2)}) + \dots \right] = \hat{H}_{\mathrm{A}}^{\mathrm{RP}} + \hat{H}_{\mathrm{B}}^{\mathrm{RP}} + \dots  \approx \hat{H}_{\mathrm{A}}^{\mathrm{RP}} + \hat{H}_{\mathrm{B}}^{\mathrm{RP}}
\end{split}
\label{eq:dipole}
\end{equation}
\hspace{0.25cm}
\end{widetext}

in which $\mu_0$ is the vacuum permeability, $s$ is the radical pair separation distance, and  $\vec{s}$ is the connection vector between the spatial positions of the coupled spins. In Eq. \eqref{eq:zeeman} and Eq. \eqref{eq:dipole}, $\vec{\hat{S}}^{(i)}=(\hat{S}_x^{(i)}, \hat{S}_y^{(i)}, \hat{S}_z^{(i)})$ are the spin operators of the radical pair's electron spins. We neglect contributions due to the exchange interaction, assuming sufficiently large radical-pair separations. 

Sensing with NV$^-$ centers commonly operates at magnetic field strengths of several tens of mili-Tesla (mT) such that the high-field approximation holds (\textbf{Fig. \ref{fig:1}c}). Therefore, we truncate $\hat{H}_{\mathrm{dip}}^{\mathrm{RP}}$ and only account for secular $\hat{H}_{\mathrm{A}}^{\mathrm{RP}}$ and pseudosecular $\hat{H}_{\mathrm{B}}^{\mathrm{RP}}$ contributions. Furthermore, in the high-field approximation, asymmetries in couplings due to other electron or nuclear spins can effectively be treated as additional differences in $g_1$ and $g_2$, and are therefore not included separately.

Differences in the $g$-factors of the radical pairs' electron spins govern the interconversion of $\ket{S}$ and $\ket{T_0}$ states. While essential for the radical pair mechanism, these dynamics do not affect the polarization $p$ of the radical pair. Rather, the  pseudosecular component of the dipolar coupling, $\hat{H}_{\mathrm{B}}^{\mathrm{RP}}$ is of prime importance. This coupling induces oscillations between the oppositely polarized $\ket{P_\mathrm{S}}$ and $\ket{P_\mathrm{R}}$ states, thereby reversing the polarization $p$ in a periodic fashion. The efficiency of these dipolar flip-flop dynamics relies on the degeneracy of the coupled spins. Thus, under the pseudosecular dipolar coupling, evolution can in principle be mitigated by differences of the radicals' $g$-factors (Appendix Sec. \ref{sec:appendA}). However, as typical dipolar couplings of D--B--A radical pairs are on the order of several MHz, significant \textit{g}-factor differences that exceed several thousands of ppm are required when sensing in the regime of $\sim 10^1\,\si{\milli\tesla}$ in order to substantially suppress evolution under the pseudosecular dipolar coupling (\textbf{Fig. \ref{fig:1}c}).

Importantly, the dipolar flip-flop hinders sensing of $p$ in all measurement schemes where the dipolar coupling strength within the radical pair exceeds their coupling to a sensor spin. Independent of the initially prepared polarization $p(0)$, the time-averaged polarization $\overline{p} = \int_t p(t) \mathrm{d}t$ vanishes.   Sensing protocols with the NV$^-$ fall into this category since typical radical separations are $<3$ nm while accessible depths of shallow NV$^-$ centers with sufficient sensing properties lie between $\sim$ 3 and \SI{10}{nm} (\textbf{Fig. \ref{fig:1}c}).\cite{Abendroth2022} Consequently, any sensing protocol that employs a weakly coupled external sensor spin to measure $p$ requires active decoupling of the dipolar interaction within the radical pair. 

\begin{figure*}
\includegraphics[width=1.0\textwidth]{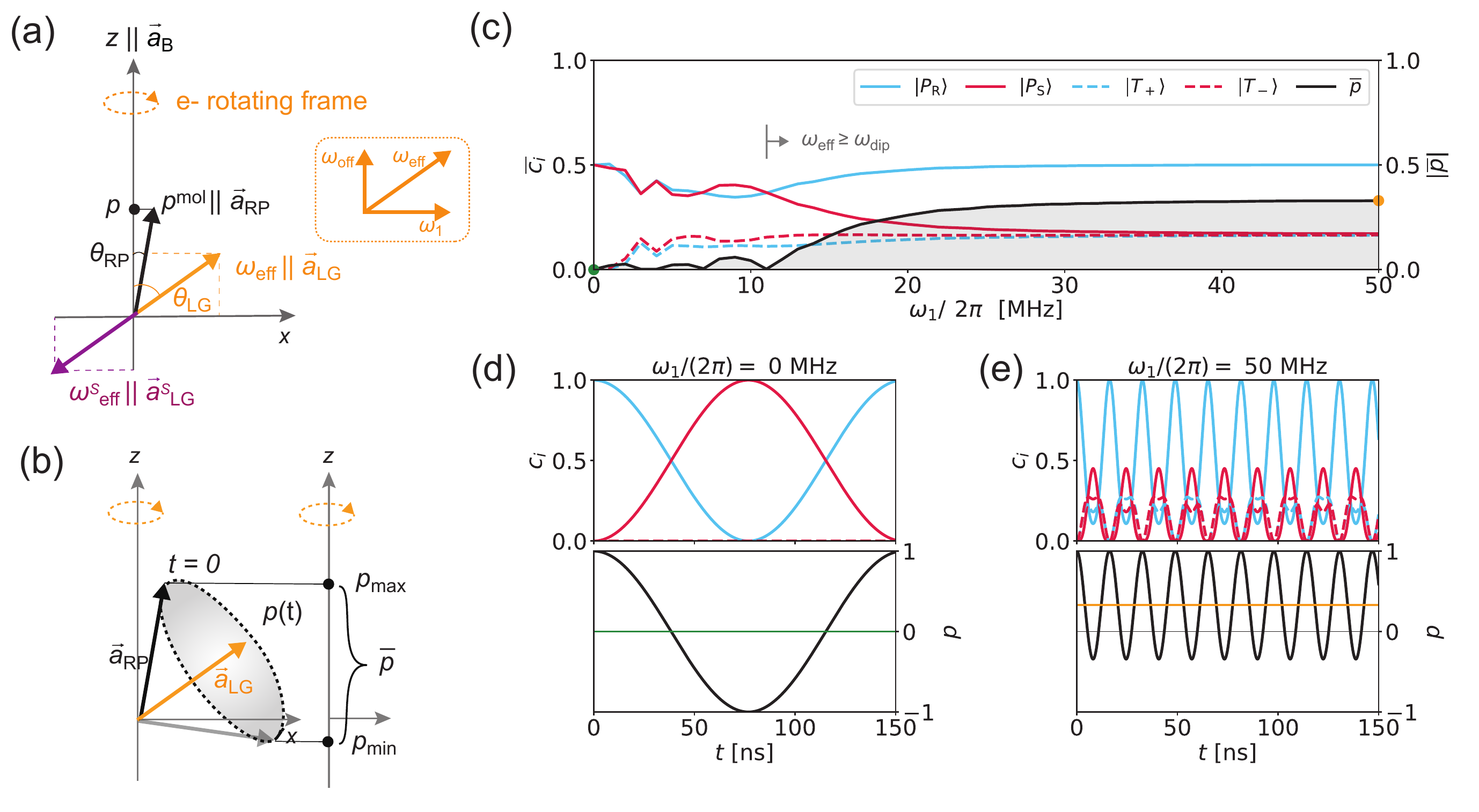}
\caption{\textbf{(a)} Lee-Goldburg (LG) and frequency-switched Lee Goldburg (FSLG) resonance conditions in the rotating frame of the electron spin. The frequency offset $\omega_{\mathrm{off}}$ of the microwave irradiation with amplitude $\omega_1$ is chosen such that the resulting effective field with amplitude $\omega_{\mathrm{eff}}$ and direction $\vec{a}_{\mathrm{LG}}$ stands at the magic angle $\theta_{\mathrm{LG}}=$\SI{54.74}{\degree} to the magnetic field direction $\vec{a}_{\mathrm{B}}$. In FSLG, the direction of $\vec{a}_{\mathrm{LG}}$ is periodically reversed into $\vec{a}^{\mathrm{S}}_{\mathrm{LG}}$, which effectively reverses the sense of rotation. \textbf{(b)} At sufficient decoupling strengths, the initial polarization $p(0)$, which can be visualized as a vector parallel to the molecular axis $\vec{a}_{\mathrm{RP}}$, evolves on a cone around $\vec{a}_{\mathrm{LG}}$.  The time-averaged polarization $\overline{p}$ can be analytically calculated by taking into account the projection of all vector positions of the trajectory $p(t)$ onto the $z$-axis of the laboratory frame of reference. \textbf{(c)} Time-averaged state occupations $\overline{c}_i = \overline{\braket{\phi_i | \rho | \phi_i}}$ for $\ket{P_\mathrm{R}}$, $\ket{P_\mathrm{S}}$, $\ket{T_+}$, and $\ket{T_-}$ states and time-averaged polarization $\overline{p}$ retrieved from explicit spin-dynamic simulations for a single radical pair with $s=\SI{2}{nm}$ and $p^{\mathrm{mol}}=$1 over a total duration of $\SI{1}{\mu s}$ (see Appendix Sec. \ref{sec:appendE} for all simulation parameters). \textbf{(d)} Time-evolution of state occupations $c_i$ and polarization $p$ without decoupling ($\omega_1=\SI{0}{MHz}$). The pseudosecular dipolar coupling continuously interconverts $\ket{P_\mathrm{R}}$ and $\ket{P_\mathrm{S}}$ state occupations at a  frequency $\omega_{\mathrm{dip}} \approx \SI{12}{MHz}$, resulting in $\overline{p}=$ 0 (green line). \textbf{(e)}  At sufficient decoupling strengths ($\omega_{\mathrm{eff}} \geq \omega_{\mathrm{dip}}$), the analytically expected $\overline{p}$ of 0.33 (orange line) is approached as the initial polarization evolves around the LG-axis instead of being periodically reversed.}
\label{fig:LGdecoupling}
\end{figure*}

Lee-Goldburg (LG) decoupling, commonly used for homonuclear decoupling in solid-state NMR spectroscopy of static samples,\cite{Lee1965} can be used to suppress the flip-flop dynamics. Analogous to spatial decoupling of dipolar interactions with magic-angle spinning (MAS), where the sample is rotated at the so-called magic angle of $\SI{54.74}{\degree}$ relative to the magnetic field axis,\cite{Bockmann2015} LG decoupling instead achieves decoupling in \textit{spin space }by creating an effective field along the magic angle $\theta_\mathrm{LG}$ in the rotating frame of the electron spins. Off-resonance transverse irradiation is applied whose frequency offset ($\omega_{\mathrm{off}}$) and amplitude ($\omega_1$) are chosen to fulfill
\begin{equation}
\theta_{\mathrm{LG}} = \arctan\frac{\omega_1}{\omega_{\mathrm{off}}} \approx \SI{54.74}{\degree}
\end{equation}
such that $1-3\cos^2{\theta_{\mathrm{RP}}} = 0$. When the decoupling strength exceeds the strength of the dipolar interaction ($\omega_{\mathrm{eff}} = \sqrt{\omega_1^2 + \omega_{\mathrm{off}}^2} \gg \omega_{\mathrm{dip}}$), the spins of the radical pairs are effectively decoupled and their trajectory is a precession around the effective LG axis $\vec{a}_{\mathrm{LG}}$ (\textbf{Fig. 2a}). In this case, $\overline{p}$ can be calculated analytically over any multiple of LG evolution periods as
\begin{equation}
\overline{p} = \frac{1}{\sqrt{3}}(\vec{a}_\mathrm{RP}\cdot\vec{a}_{\mathrm{LG}})p_{\mathrm{CISS}},
\label{eq:Gamma}
\end{equation}
which comprises two scaling contributions to $p_{\mathrm{CISS}}$. First, polarization in the molecular axis $\vec{a}_\mathrm{RP}$ is projected onto the axis of the effective LG-field $\vec{a}_{\mathrm{LG}}$. A second projection onto the magnetic field direction results in the factor $\cos{\theta_\mathrm{LG}} = 1/\sqrt{3}$. 
 (\textbf{Fig. \ref{fig:LGdecoupling}b}). Importantly, LG decoupling preserves the relative signs of spin polarization, which enables enantiomer discrimination and does not induce any time-averaged polarization when starting from a non-polarized state.

Analysis of spin-state evolution reveals that the $\ket{T_+}$ and $\ket{T_-}$ states are partially populated as a result of LG decoupling (Appendix Sec. \ref{sec:appendB}). While occupied equally in the limit of sufficiently strong decoupling, asymmetries in occupations of $\ket{T_+}$ and  $\ket{T_-}$ are present in intermediate regimes of incomplete decoupling. These asymmetries induce a net-polarization in the double quantum subspace that complicates sensing of $p$. To remove these asymmetries, we employ frequency-switched Lee-Goldburg (FSLG) decoupling,\cite{Bielecki1989, Levitt1993} where the sense of rotation is reversed after every full $2\pi$ rotation around the magic angle axis. Compared to simplified LG decoupling that uses continuous-wave irradiation with constant frequency and phase, FSLG provides an experimentally tractable route that reduces the required decoupling amplitudes. Explicit spin dynamics simulations for a single isolated radical pair with $s=$ \SI{2}{nm} in an initial state $\ket{P_\mathrm{R}}$ aligned with $\vec{B}$ ($p=p_{\mathrm{CISS}}=1$) reveal that $\overline{p}=0$ when $\omega_1=$  \SI{0}{MHz} or when decoupling strength does not exceed the dipolar interaction strength (\textbf{Fig. \ref{fig:LGdecoupling}c,d}). Alternatively, with increasing $\omega_1$, under sufficient decoupling the polarization reaches a maximum value of $\overline{p}=0.33$ (\textbf{Fig. \ref{fig:LGdecoupling}e}), consistent with the analytically expected scaling (Eq. \eqref{eq:Gamma}).

\begin{figure}
\includegraphics[width=0.48\textwidth]{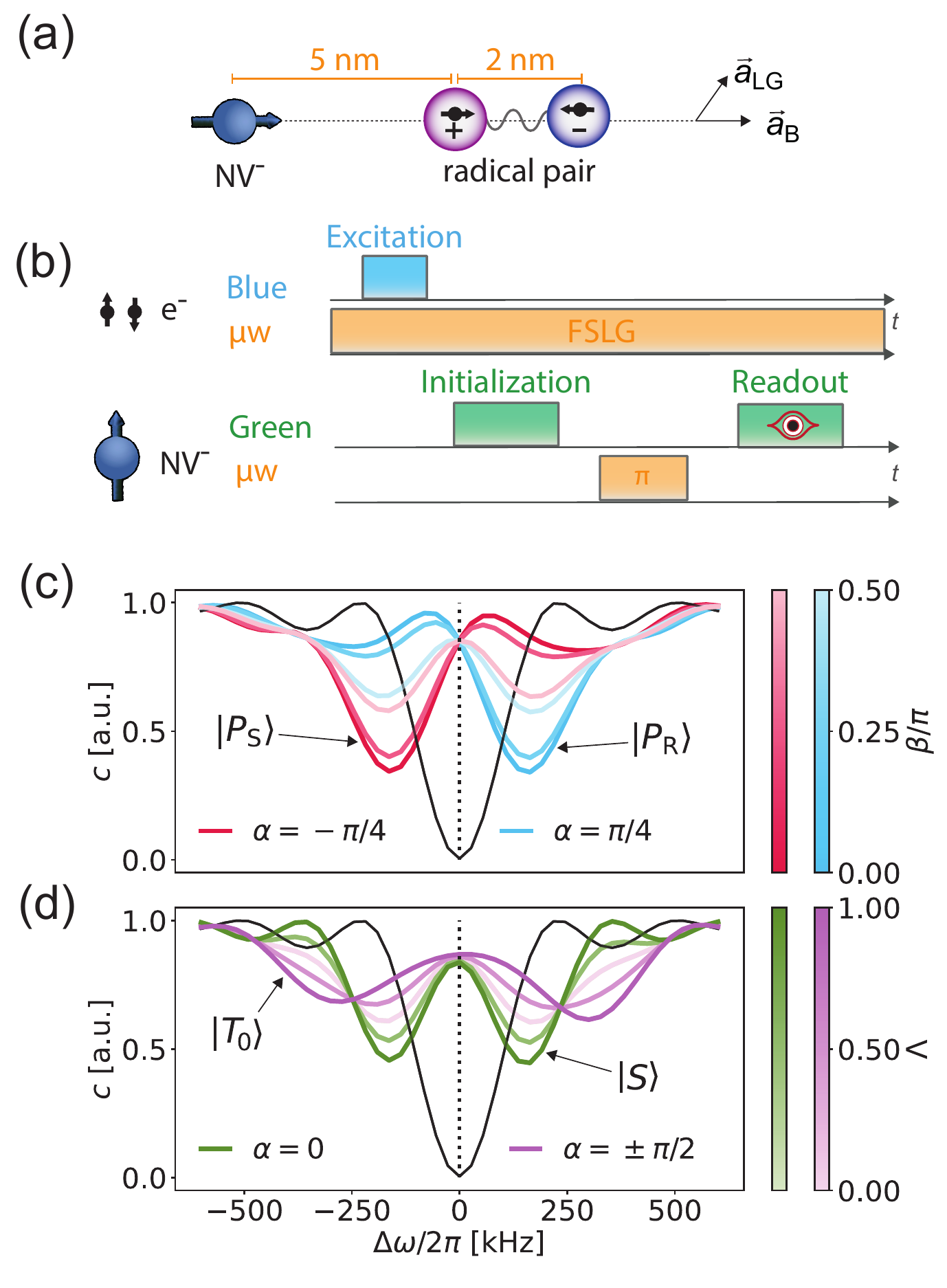}
\caption{ Explicit spin simulations of an ODMR experiment for NV$^-$ sensing of a single radical pair. \textbf{(a)}  Schematic of the geometry of the simulated three-spin system, indicating distances between the spins and directions of magnetic ($\vec{a}_\mathrm{B}$) and LG decoupling ($\vec{a}_{\mathrm{LG}}$) fields. \textbf{(b)} Pulsed ODMR sequence for the simulated experiment. Blue laser excitation could be used to induce charge separation in the D--B--A molecule and generate a radical pair. FSLG decoupling supplied by $\upmu$w irradiation on the electron spins of the radical pair is applied throughout excitation and successive detection with the NV$^-$ center. A green laser is used for initialization and fluorescent readout of the NV$^-$ and $\upmu$w irradiation in the form of a $\pi$-pulse is applied to reveal resonant frequencies. \textbf{(c,d)} Simulated ODMR spectra of an NV$^-$ without (black lines) and with (colored lines) coupling to a single radical pair initialized in different initial states $\rho_0(\alpha, \beta, \Lambda)$ as indicated in the color and transparencies of the respective spectra. The optical contrast $c$ is plotted against the frequency difference $\Delta\omega$ to the resonance frequency of the NV$^{-}$ center's $\ket{0} \rightarrow \ket{-1}$ transition in the absence of any dipolar couplings. \textbf{(c)} If $\alpha = \frac{\pi}{4}$ (blue) or $-\frac{\pi}{4}$ (red), the initial states are oppositely polarized according to $p = \pm \cos(\beta)$, respectively. If $\beta = 0$, fully polarized states $\ket{P_\mathrm{R}}, \ket{P_{\mathrm{S}}}$ are obtained for opposite handedness of the bridge ($R, S$) that can be distinguished. If $\beta = \pm \frac{\pi}{2}$, enantioselectivity of the sensing protocol vanishes and only line-shape asymmetries due to decoupling artifacts remain. \textbf{(d)} If $\alpha = 0$ (green) or $\pm\frac{\pi}{2}$ (lilac), the initial states are unpolarized, $p = 0$, independent of the value of $\beta$, and symmetric line splittings are observed instead of a frequency shift. If coherent states are initialized ($\Lambda = 1$) we recover $\ket{S}$ and $\ket{T_0}$ states, respectively, which can be distinguished under FSLG decoupling. In the limit of $\Lambda=0$, they become indistinguishable. }
\label{fig:singlemolecule}
\end{figure}

We now turn to the NV$^-$ as a local probe to distinguish initial states in a radical pair subjected to FSLG decoupling. The ODMR sensing relies on the electronic ground state of the NV$^-$, with "bright" $m_\mathrm{T}=0$ and "dark" $m_\mathrm{T}=\pm1$ spin sublevels separated by a zero-field splitting of $D=\SI{2.87}{GHz}$; the degenerate $m_\mathrm{T}=\pm1$ energy levels can be split with a magnetic bias field along the defect axis. Following off-resonant irradiation with a green laser to initialize the NV$^-$ in the $m_\mathrm{T}=0$ state, transitions between $\ket{0}$ and $\ket{\pm 1}$ sub-levels are driven by microwave irradiation. Readout is performed by monitoring fluorescence upon application of a second laser pulse. Under appropriate sensing protocols with the NV$^-$, the magnetic moment of nearby spins can be detected as a shift in phase\cite{Staudacher2013} or in frequency\cite{Mamin2013} of the defect's resonance condition. A full description of the electronic ground state Hamiltonian and how this state couples to or is perturbed by the environment can be found elsewhere.\cite{Janitz2022} Here, we limit the discussion to the detection of CISS \textit{via} a shift of the NV$^-$ resonance frequency as a result of DC magnetic fields from a proximal spin-polarized radical pair. Such sensing of longitudinal polarization of external spins is mediated by the secular term of the dipolar coupling between the NV$^-$ and the i$^{\mathrm{th}}$ external spin
\begin{equation}
\hat{H}_{\mathrm{A}}^{(i)} =  \frac{\mu_0 \hbar \gamma_{\mathrm{NV}}\gamma_i}{4 \pi}\frac{1}{r_i^3} ( 1 - 3 \cos^2{\theta_i}) \hat{T}_{z}\hat{S}_{z}^{(i)}
\label{eq:dc_sensing_coupling}
\end{equation}
where $\gamma_{\mathrm{NV}}$ and $\gamma_{i}$ are the gyromagnetic ratios of the NV$^-$  and the i$^{\mathrm{th}}$ external spin, $r_i$ is their distance, $\theta_i$ is the angle between their connection axis and the external magnetic field and $\vec{\hat{T}} = (\hat{T}_x, \hat{T}_y, \hat{T_z})$ are the spin operators for the NV$^-$ electron spin. For $\braket{\hat{S}_{z}^{(i)}} = \pm \frac{1}{2}$ a target electron spin induces resonance frequency shifts of $\Delta \omega_{\mathrm{A}}^{(i)} = \pm \frac{\mu_0 \hbar \gamma_{\mathrm{NV}}\gamma_i}{8 \pi}\frac{1}{r_i^3} ( 1 - 3 \cos^2{\theta_i})$ where the global sign is reversed for the $\ket{0} \rightarrow \ket{+1}$ and $\ket{0}\rightarrow \ket{-1}$ transitions. When the NV$^-$ couples to two electron spins of a radical pair, four distinct frequency shifts 
\begin{equation}
\Delta\omega_{\mathrm{A}} =  \Delta\omega_{\mathrm{A}}^{(1)} +  \Delta\omega_{\mathrm{A}}^{(2)}
\label{eq:fullshiftsingle}
\end{equation}
are possible accounting for all combinations of $\braket{\hat{S}_{z}^{(1)}} = \pm \frac{1}{2}$ and $\braket{\hat{S}_{z}^{(2)}} = \pm \frac{1}{2}$ that are the eigenstates of the Zeeman basis. The ODMR spectrum becomes a superposition of these frequencies whose relative intensities are determined by the time-averaged populations $\overline{c_i}$ of the Zeeman basis states resulting from the evolution of $\rho_0$ during the sensing protocol. The time-averaged polarization $\overline{p}$ reveals itself as an asymmetry in the observed lineshape. If no dipolar decoupling is applied, the shifts for $\ket{P_{\mathrm{S}}}$ and $\ket{P_{\mathrm{R}}}$ states average to zero in cases where the inverse of the dipolar coupling frequency exceeds the duration of the sensing protocol, independent of the initial spin state. Under LG decoupling, this averaging is prohibited, and the magnitudes of the induced frequency shifts (Eq. \eqref{eq:fullshiftsingle}) are merely scaled to  $\Delta \omega_{\mathrm{A}}^ {\mathrm{LG}} = \Delta \omega_{\mathrm{A}}/\sqrt{3}$ resulting from the transformation of the LG axis into the laboratory frame.

Calculated ODMR spectra from explicit spin-dynamics simulations are shown in \textbf{Fig. 3} for an NV$^-$ spin positioned 5 nm away from the closest electron in a collinearly aligned radical pair spin system. Here, the magnetic field is aligned along the three-spin axis and $\gamma_1 = \gamma_2 = \gamma_e$ (see Appendix Sec. \ref{sec:appendE} for simulation details). Under FSLG decoupling ($\omega_1/2\pi=\SI{50}{MHz}$), radical pairs initially in $\ket{P_{\mathrm{S}}}$ or $\ket{P_{\mathrm{R}}}$ states can be distinguished from each other and from $\ket{S}$ and $\ket{T_0}$ states. The observed ODMR lineshapes can be understood based on the time-averaged state occupations $\overline{c}_i$ under FSLG decoupling for the respective initial spin states (\textbf{Fig. \ref{fig:LGdecoupling}}, Appendix Sec. \ref{sec:appendB}). Note, the simulation results also hold when $\gamma_1\neq\gamma_2$, which results in only minor perturbation of the lineshapes. In addition, changes in orientation of the radical pair with respect to the NV$^-$ (\textit{e.g.}, due to molecular motion) either between successive repetitions of the experiment or during a single experiment would result in line broadening, but not alter the basic principle of the sensing experiment.

If CISS does not induce any polarization in the initial state (\textit{e.g.}, $\beta = \pm \frac{\pi}{2}$ and therefore $p=0$), asymmetry is lost in the ODMR spectrum (\textbf{Fig. \ref{fig:singlemolecule}c}). In this case, alternative sensing protocols would be necessary to probe the mixing angle $\alpha$ between $\ket{S}$ and $\ket{T_0}$ that is possibly influenced by CISS.\cite{Fay2021} For finite $p$, however, discrimination of $\ket{P_{\mathrm{S}}}$ and $\ket{P_{\mathrm{R}}}$ states is possible independent of the extent of coherence, as $p$ is independent of $\Lambda$ (Eq. \eqref{eq:polarization}). Still, $\Lambda$ can influence the observed lineshape for states with $|p|<1$ in a symmetric fashion (\textbf{Fig. \ref{fig:singlemolecule}d}).

\begin{figure*}
\includegraphics[width=1.0\textwidth]{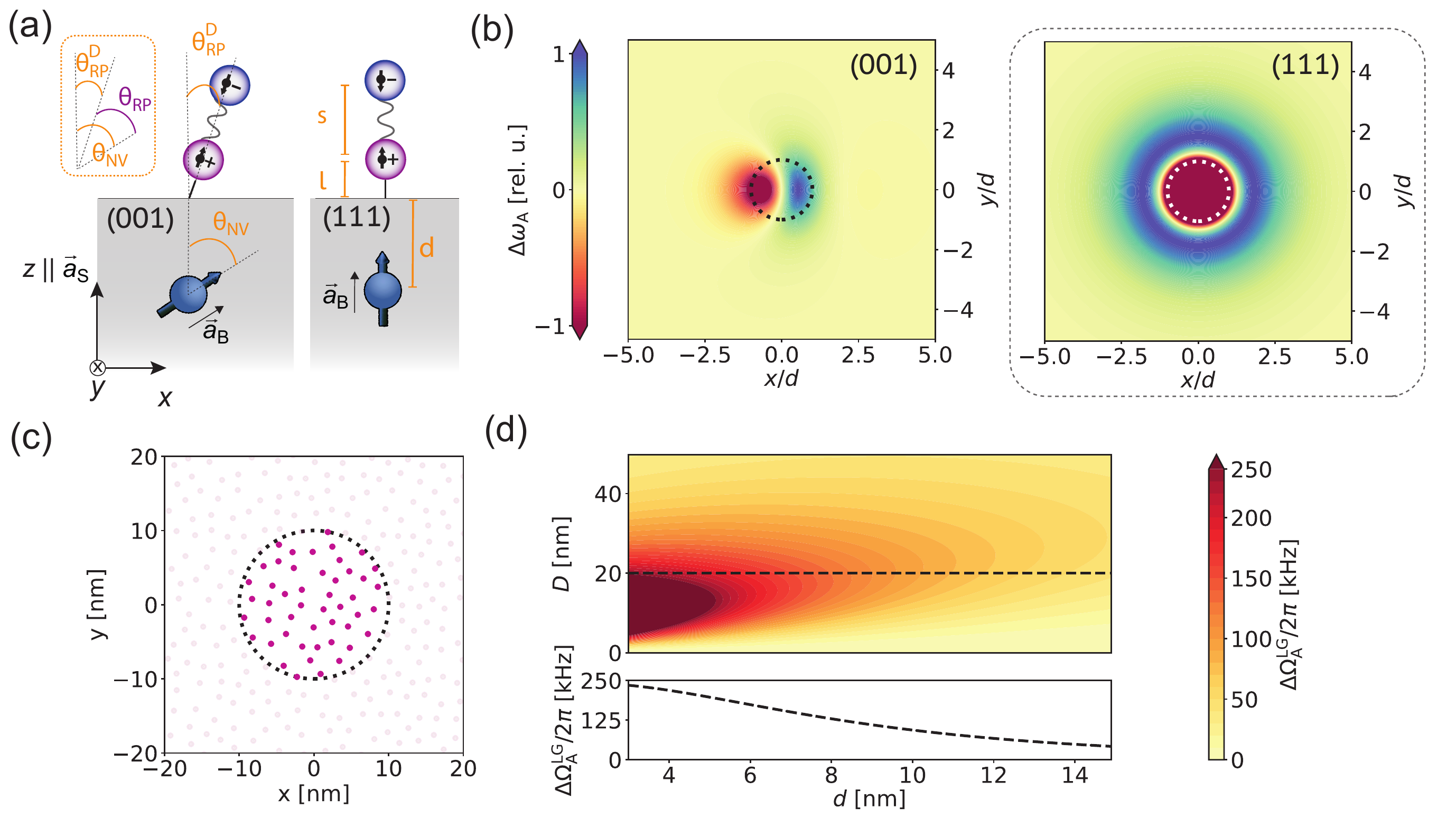}
\caption{\textbf{(a)} Visualization of relevant geometries for simulated monolayers of D--B--A molecules with linker length \textit{l} on diamond (001) or (111) surfaces. \textbf{(b)} Geometric sensitivity profiles for (001)-terminated diamond (\textit{left}) and (111)-terminated diamond (\textit{right}). For varying positions of molecules on the diamond surface (normalized to NV$^-$ depth $d$) $x/d$, $y/d$ the relative (normalized) induced frequency shift on the NV$^-$ $\Delta\omega_{\mathrm{A}}^{\mathrm{rel}}$ is given. \textbf{(c)} Visualization of one possible surface geometry obtained in a Monte Carlo type simulation of anchoring positions of D--B--A molecules assuming a density $\rho_{\mathrm{anchor}}=\SI{5}{anchors/nm^2}$ and a molecular footprint $D= $ \SI{2}{nm} (See Appendix Sec. \ref{sec:appendE}). The dotted black circle indicates masking, \textit{i.e.}, by molecular patterning with $D=$ \SI{20}{nm} to achieve enantioselectivity on the (111) surface assuming an underlying NV$^-$ centered at $(0,0)$ with depth \textit{d}. \textbf{(d)} Expected frequency shifts $\Delta\Omega_{\mathrm{A}}^{\mathrm{LG}}$ for oriented monolayers of D--B--A molecules supporting radical pairs in $\ket{P_\mathrm{S}}$ or $\ket{P_\mathrm{R}}$ states assuming molecular densities $\rho= \SI{0.15}{molecules/nm^2}$ for varying $d$ and $D$.}
\label{fig:monolayers}
\end{figure*}

An experimental realization of the proposed sensing scheme with single NV$^-$ centers also necessitates consideration of the diamond crystal surface. The nitrogen impurity and adjacent vacancy of the NV$^-$ are projected along the crystallographic $\braket{111}$ direction of the diamond lattice. This direction defines the direction of the magnetic bias field ($\vec{a}_{\mathrm{B}}$) which is aligned with this axis. Considering a (001) surface, the NV$^-$ occupies one of four possible orientations each at $\theta_{\mathrm{NV}}= $ \SI{54.74}{\degree} relative to the surface normal. For the (111) surface, one possible orientation is aligned parallel to the surface normal ($\theta_{\mathrm{NV}}= $ \SI{0}{\degree}), while the other three axes are projected at $\theta_{\mathrm{NV}}=$ \SI{109.47}{\degree}. Here, we only consider $\theta_{\mathrm{NV}}=$ \SI{0}{\degree} for (111) surfaces. Critically, the choice of (001) \textit{versus} (111) surfaces influences the viability of the sensing scheme in two ways: 
\begin{enumerate}
\item[\textbf{1}] The transformation of the spin state of the radical pair from the molecular frame into the laboratory frame is governed by a rotation of the underlying Zeeman basis by $\theta_{\mathrm{RP}}$ (Eq. 4). This angle can be calculated from  $\theta_{\mathrm{RP}} = \theta_{\mathrm{NV}}-\theta^{\mathrm{D}}_{\mathrm{RP}}$, where  $\theta^{\mathrm{D}}_{\mathrm{RP}}$ is the angle of the molecular axis $\vec{a}_{\mathrm{RP}}$ relative to the diamond surface normal $\vec{a}_{\mathrm{S}}$ (\textbf{Fig. 4a}). 
\item[\textbf{2}] The geometric sensitivity profile of the NV$^-$ for longitudinal polarization, \textit{i.e.}, the spatial part of $\hat{H}_{\mathrm{A}}^{(i)}$, depends on $\theta_{NV}$.\cite{Bruckmaier2021} For both (001) and (111) surfaces, the sign of $\Delta\omega_{\mathrm{A}}$  therefore varies across the surface (\textbf{Fig. 4b}). As a result, in the case of single molecule adsorbed on the surface, the sign of $\Delta\omega_{\mathrm{A}}$ is not unique to $\ket{P_\mathrm{S}}$ or $\ket{P_\mathrm{R}}$ and depends on the molecule's position with respect to the NV$^-$. If a homogeneous monolayer of molecules lie within the sensing volume, the total measurable shift would vanish due to cancelling effects. Enantioselectivity in these systems could be recovered, however, by nanostructuring of the diamond, inhomogeneous monolayers, or molecular surface patterning to avoid cancellation of $\Delta\omega_{\mathrm{A}}$ by regions of opposite sign. 
\end{enumerate}

Assuming molecules are adsorbed and oriented perpendicular to the surface ($\theta^{\mathrm{D}}_{\mathrm{RP}}=0$), $p_{\mathrm{CISS}}$ is maximally preserved in the laboratory frame for (111) surfaces. Further, the geometric sensitivity profile of the (111) surface possesses radial symmetry and is better suited for masking regions of positive or negative $\Delta\omega_{\mathrm{A}}$ than (001)-terminated diamond. For example, concentric patterning of molecules over NV$^-$ centers could be realized experimentally by proper mask alignment during both the nitrogen implantation step for defect formation\cite{Toyli2010} as well as the chemical patterning step.

Adopting the (111) surface as the preferred sensing platform, we evaluated the result of a sensing experiment of spin-polarized radical pairs in patterned monolayers of D--B--A molecules. Note, if the NV$^-$ center couples to multiple radical pairs the resulting frequency spectrum is expected to broaden. Still, the characteristic asymmetry due to CISS would remain as long as a (partial) polarization is present in the majority of the molecules. The resulting intensity maximum of the frequency spectrum lies at the sum of the shifts for the spin-polarized states of the individual molecules, $\Delta\Omega_{\mathrm{A
}}^{\mathrm{LG}}$. We analytically evaluated the expected intensity maxima $\Delta\Omega_{\mathrm{A
}}^{\mathrm{LG}}$ for various NV depths $d$ and mask diameter $D$ using the analytical expression derived in Appendix Sec. \ref{sec:appendC}. We chose a conservative molecular surface density of \SI{0.15}{molecules/nm^2} based on Monte Carlo type simulations of possible surface configurations (\textbf{Fig. \ref{fig:monolayers}c} and Appendix Sec. \ref{sec:appendD}).
As shown in \textbf{Fig. \ref{fig:monolayers}d}, shifts of several hundred kHz are expected for radical pairs in $\ket{P_\mathrm{S}}$ or $\ket{P_\mathrm{R}}$ states, which is well within the detection sensitivity using established NV$^-$ sensing protocols.\cite{Janitz2022} In particular, using $D=$ \SI{20} {nm}, achievable using standard electron-beam lithography methods with high-resolution resists, maximum shifts of $\Delta\Omega_{\mathrm{A}}^{\mathrm{LG}}/2\pi>100$ kHz are possible when the NV$^-$ is $\sim$ \SI{10}{nm} or closer to the surface.

In conclusion, we developed a sensing scheme using single NV$^-$ centers in diamond as quantum sensors to probe the hypothesized spin polarization of initial states in radical pairs that results from charge transfer across a chiral bridge. Using realistic experimental parameters, we addressed both single-molecule and ensemble-type sensing schemes where multiple molecules are probed by one sensor. Although sensing the polarization is the first main interest, measurements using a quantum sensor are not limited to detection of polarized states. With the rise of quantum information processing, numerous protocols for quantum state tomography (QST) \cite{Cramer2010,Lanyon2017,Torlai2018} have been proposed for experimental elucidation of multi-qubit entangled states. Mapping out the full density matrix by means of QST will be a major step toward the theoretical understanding of the CISS effect. In essence, this requires the measurement of a complete set of observables whose expectation values determine the quantum state fully. Although, in many practical situations a smaller number of parameters ($\alpha,\beta,\Lambda$) are sufficient to describe the state. By probing these parameters with appropriate sensing protocols, it will be possible to reconstruct the initial state after the CISS effect has occurred, including states that have no net polarization.
\pagebreak
\section*{AUTHOR CONTRIBUTIONS}
\textbf{Laura A. Völker:} Conceptualization (equal);  Formal analysis (equal); Investigation (equal); Methodology (equal); Validation (equal); Visualization (lead); Writing -- original draft (equal); Writing -- review \& editing (equal). \textbf{Konstantin Herb:} Conceptualization (equal); Formal analysis (equal); Investigation (equal); Methodology (equal); Validation (equal);  Writing -- original draft (equal); Writing -- review \& editing (equal). \textbf{Erika Janitz:} Validation (supporting); Writing -- review \& editing (equal). \textbf{Christian L. Degen:} Funding acquisition (supporting); Validation (equal); Supervision (supporting); Writing -- review \& editing (equal). \textbf{John M. Abendroth:} Conceptualization (equal); Funding acquisition (lead); Project administration (lead), Supervision (lead); Validation (equal); Writing -- original draft (equal); Writing -- review \& editing (equal).

\begin{acknowledgments}
We gratefully acknowledge the Swiss National Science Foundation (SNSF) Ambizione Grant, "Quantum Coherence in Mirror-Image Molecules" [PZ00P2-201590] and SNSF  NCCR QSIT, a National Centre of Competence in Research in Quantum Science and Technology, Grant No. 51NF40-185902 for support of this worrk. J.M.A. also acknowledges funding from an ETH Zurich Career Seed Grant and E.J. acknowledges support from a Natural Sciences and Engineering Research Council of Canada (NSERC) postdoctoral fellowship (PDF-558200-2021).
\end{acknowledgments}

%\section*{Data Availability Statement}
%The data that support the findings of this study are available from the corresponding author upon reasonable request. 

\appendix

\renewcommand\thefigure{A\arabic{figure}}
\renewcommand\thetable{A\arabic{table}}

\setcounter{figure}{0}

\section{Spin States and Dynamics of Radical Pairs - An Extended Overview}\label{sec:appendA}

\begin{figure*}
\includegraphics[width=1.0\textwidth]{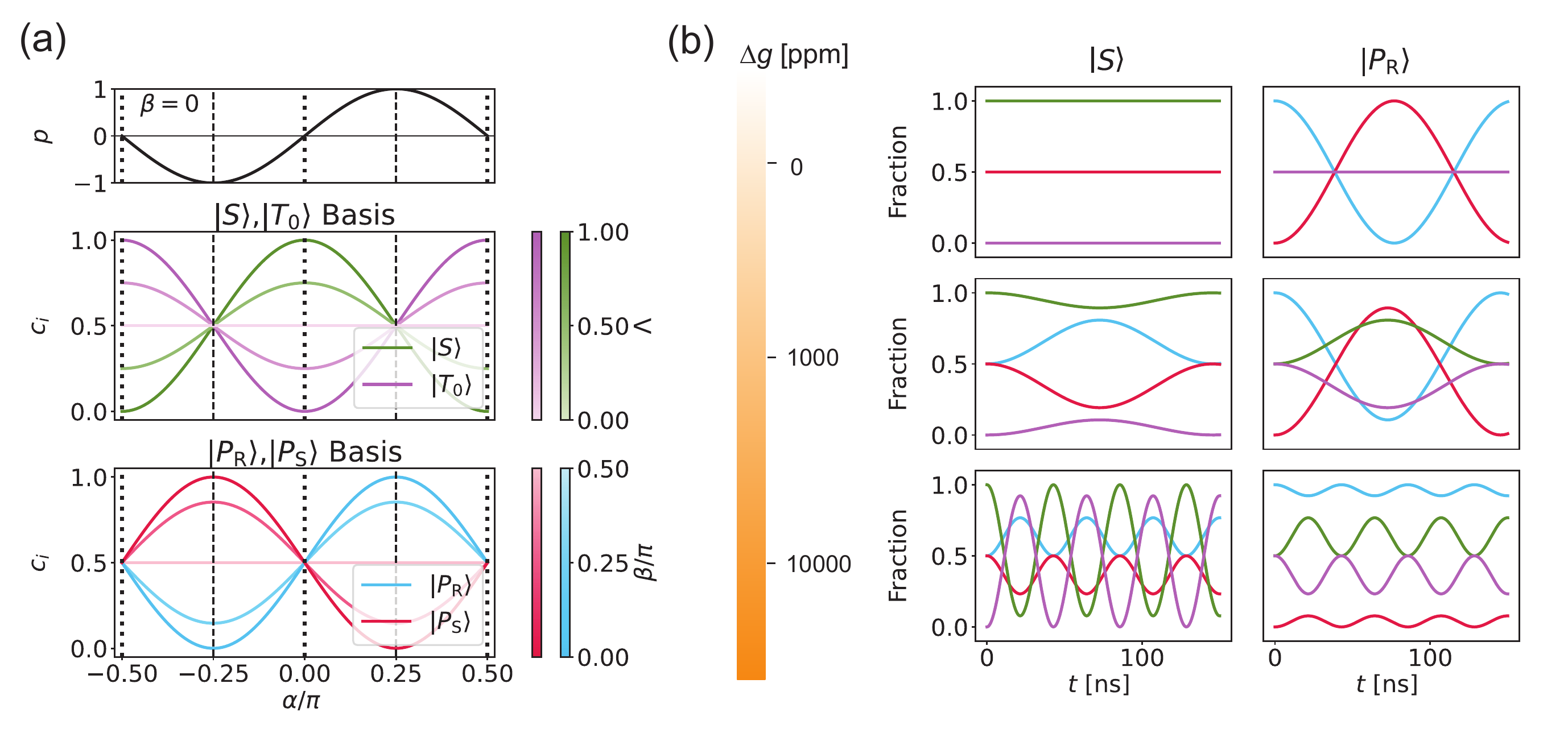}
\caption{ \textbf{(a)} Spin states of spin-correlated radical pairs. Relative state occupations $c_i$ are evaluated in a singlet-triplet ($\ket{S}, \ket{T_0}$) basis as well as a polarization ($\ket{P_{\mathrm{R}}}, \ket{P_{\mathrm{S}}}$) basis as defined in Eq. \eqref{eq:initial_state} and Eq. \eqref{eq:initial_state_alt} for the entire range of possible $\alpha$. The influence of $\Lambda$ is shown in the $\ket{S}, \ket{T_0}$-basis (\textit{middle}). The influence of $\beta$ is visualized in the $\ket{P_{\mathrm{R}}}, \ket{P_{\mathrm{S}}}$ basis (\textit{bottom}). \textbf{(b)} Time evolution of a singlet $\ket{S}$ state ($\alpha =0$, $\Lambda = 1$) and a polarized state $\ket{P_{\mathrm{R}}}$ ($\alpha=\pi/4$, $\beta = 0$) for varying $g$-factor differences of the radical pairs electron spins. }
\label{fig:spinstates}
\end{figure*}

Recent work on radical-pair spin dynamics commonly makes use of a singlet--triplet basis ($\ket{\phi_i} = \ket{S}, \ket{T_0}$) for the zero-quantum subspace, in which the initial spin state of the spin-correlated radical pair is expressed as in Eq. \eqref{eq:initial_state}. However, an alternative basis which is more natural in the context of this work utilizes the polarized states ($\ket{\phi_i} = \ket{P_{\mathrm{R}}}, \ket{P_{\mathrm{S}}}$) which are also used as basis states for the zero-quantum subspace in Eq. \eqref{eq:initial_rho} such that
\begin{equation}
\ket{\psi_0} = \frac{\cos{\alpha} + e^{i\beta}\sin{\alpha}}{\sqrt{2}} \ket{P_{\mathrm{R}}} - \frac{\cos{\alpha} - e^{i\beta}\sin{\alpha}}{\sqrt{2}} \ket{P_{\mathrm{S}}}
\label{eq:initial_state_alt}
\end{equation}
Notably, the definitions provided in 
Eqs. \eqref{eq:initial_state} and \eqref{eq:initial_state_alt} are fully equivalent. Table A1 shows a summary of recent definitions of initial spin states of radical pair systems in the context of CISS and their associated references. In \textbf{Fig. \ref{fig:spinstates}} we visualize state occupations $c_i = \bra{\phi_i}\rho_0 \ket{\phi_i}$ in both bases for a complete set of ($\alpha, \beta, \Lambda$). While $\alpha$ determines the degree of mixing of the two states and is therefore represented equally well in both bases, the influence of $\beta$ is only visible in the $\ket{P_{\mathrm{R}}}, \ket{P_{\mathrm{S}}}$ basis and the influence of $\Lambda$ manifests itself in the $\ket{S}, \ket{T_0}$ basis. To demonstrate the influence of $g$-factor anisotropies ($g_1 = g_e-\frac{\Delta}{2}g$ and $g_2 = g_e+\frac{\Delta}{2}g$) and the dipolar coupling of the radical spins, the simulated time evolution of  i) a singlet state ($\ket{S}$) and ii) a polarized state ($\ket{P_{\mathrm{R}}}$) are shown. When $\Delta g =$ \SI{0}{ppm}, the dipolar coupling periodically interconverts $\ket{P_\mathrm{R}}$ and $\ket{P_\mathrm{S}}$ states. If significant $\Delta g =$ \SI{10000}{ppm} is present, this conversion is mitigated and spin-state dynamics are dominated by singlet--triplet interconversion.   

\begin{table*}
\caption{\label{tab:spinstates_refs} Definitions of initial spin states of spin-correlated radical pairs (SCRPs) in recent publications in the context of CISS.}
\begin{ruledtabular}
\def\arraystretch{1.5}
\begin{tabular}{lllllll}
 Reference &  $\ket{\psi_0}$ or $\rho_0$ \footnote{with variable symbols of original reference} & $\ket{\psi_0}$ or $\rho_0$ \footnote{with variable symbols of this work} & $\alpha$ 
& $\beta$ & $\Lambda$ & $p$ \\ \hline
\cite{Fay2021, Chiesa2021} &  $\ket{\psi_0} = \cos{\theta} \ket{S} + i\sin{\theta} \ket{T_0}$ & $\ket{\psi_0} = \cos{\alpha} \ket{S} + i\sin{\alpha} \ket{T_0}$ & $\in [-\frac{\pi}{2}, \frac{\pi}{2}]$  & $\frac{\pi}{2}$ & 1  & 0 \\  \hline
\cite{Fay2021b} &  $\ket{\psi_0} = \cos{\theta} \ket{S} + i\sin{\theta}e^{2iJt/\hbar} \ket{T_0}$  & $\ket{\psi_0}= \cos{\alpha} \ket{S} + e^{i\beta}\sin{\alpha} \ket{T_0}$  & $\in [-\frac{\pi}{2}, \frac{\pi}{2}]$  & $\in [-\pi, \pi]$ & 1  & $\in [-1, 1]$ \\ \hline 
\cite{Luo2021} & $\ket{\psi_0} = \cos{\frac{\chi}{2}} \ket{S} + \sin{\frac{\chi}{2}} \ket{T_0}$  & $\ket{\psi_0} = \cos{\alpha} \ket{S} + \sin{\alpha} \ket{T_0}$  & $\in [-\frac{\pi}{2}, \frac{\pi}{2}]$  & 0 & 1  & $\in [-1, 1]$ \\ \hline 
\cite{Tiwari2022} & $\ket{\psi_0} = \frac{1}{\sqrt{2}}[\sin{\frac{\chi}{2}} + \cos{\frac{\chi}{2}}]\ket{\uparrow\downarrow}$  & $\ket{\psi_0} = \cos{\alpha} \ket{S} + \sin{\alpha} \ket{T_0}$  & $\in [-\frac{\pi}{2}, \frac{\pi}{2}]$  & 0 & 1  & $\in [-1, 1]$ \\
 &  \hspace{0.6cm} $ +  \frac{1}{\sqrt{2}}[\sin{\frac{\chi}{2}} - \cos{\frac{\chi}{2}}]\ket{\downarrow\uparrow}$ & & & & & 
\\ \hline 
\cite{Chiesa2021}   & $\rho_0 = \frac{1+p}{2} \ket{\uparrow\downarrow}\bra{\uparrow\downarrow}  +  \frac{1-p}{2} \ket{\downarrow\uparrow}\bra{\downarrow\uparrow}$ & $\rho_0 = \frac{1+p}{2} \ket{P_{\mathrm{R}}}\bra{P_{\mathrm{R}}}  +  \frac{1-p}{2} \ket{P_{\mathrm{S}}}\bra{P_{\mathrm{S}}} $ & $\in [-\frac{\pi}{2}, \frac{\pi}{2}]$  & $\in [-\pi, \pi]$  & 0 & $\in [-1, 1]$ \\ \hline
this work & - & Eq. \eqref{eq:initial_state}, \eqref{eq:initial_rho} & $\in [-\frac{\pi}{2}, \frac{\pi}{2}]$  &  $\in [-\pi, \pi]$ & $\in [0, 1]$  & $\in [-1, 1]$
\end{tabular}
\end{ruledtabular}
\end{table*}

\section{Spin State Evolution under LG Decoupling and FSLG Decoupling} \label{sec:appendB}

\begin{figure*}
\includegraphics[width=1.0\textwidth]{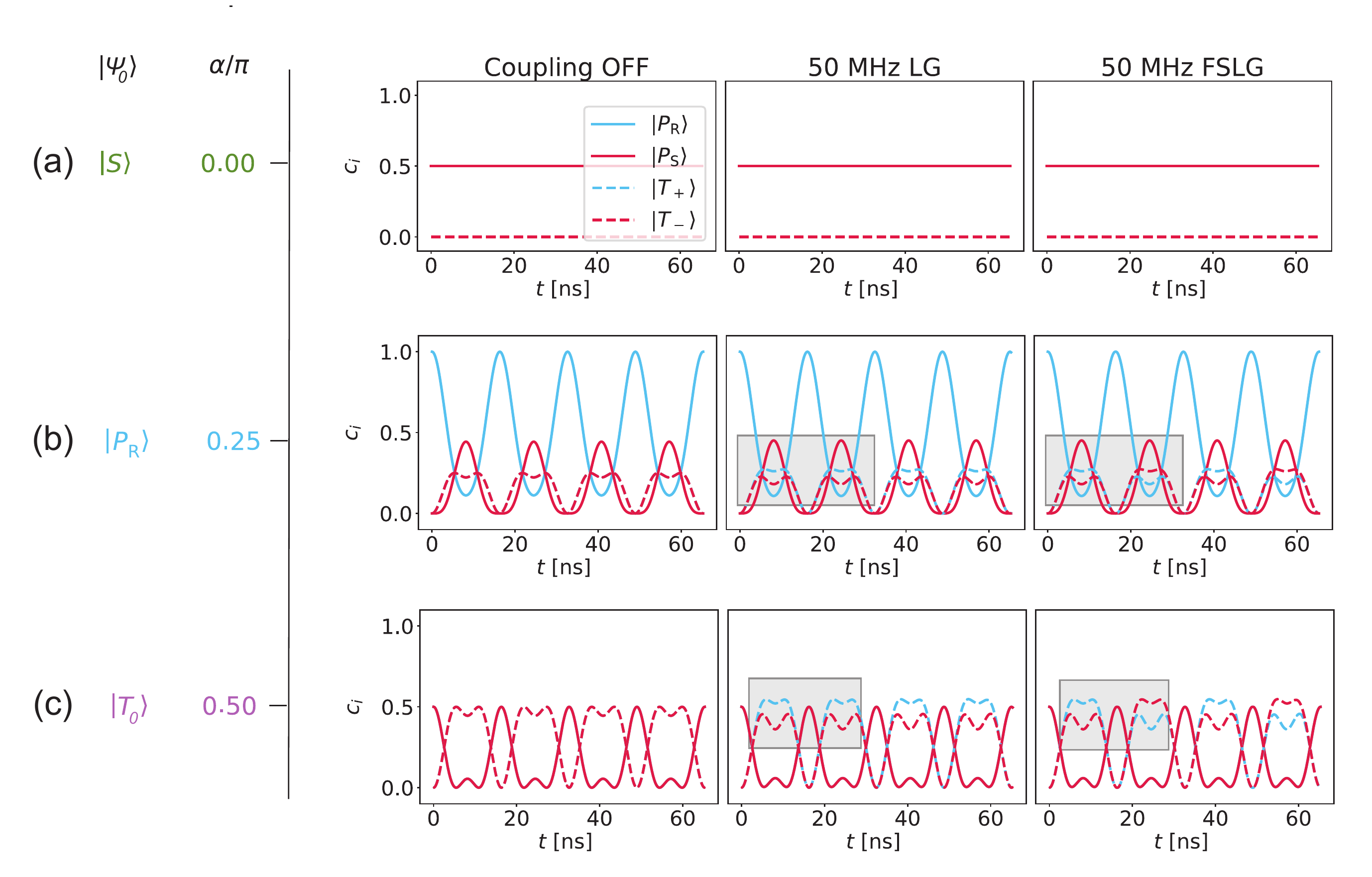}
\caption{ Rotating frame simulations for a single radical pair (for details see Sec. \ref{sec:appendB}) under LG and FSL decoupling (middle and right column) as well as evolution under LG decoupling if the dipolar coupling is neglected, which shows the limit of perfect decoupling (left column).   Evolution is shown starting from \textbf{(a)} a singlet $\ket{S}$ state, \textbf{(b)} a polarized $\ket{P_{\mathrm{R}}}$ state and \textbf{(c)} a triplet $\ket{T_0}$ state.}
\label{fig:LGextension}
\end{figure*}

The lineshapes of the simulated NV$^-$ spectra shown in \textbf{Fig. \ref{fig:singlemolecule}} can be readily understood based on  rotating frame simulations of a radical pair ($s= \SI{2}{nm}$, $\theta_{\mathrm{RP}}= 0$, $B_z= \SI{40}{mT}$, $\gamma_1 = \gamma_2 = \gamma_e$) under FSLG decoupling ($\omega_1/2\pi= \SI{50}{MHz}$) from the respective initial states as shown in \textbf{Fig. \ref{fig:LGextension}}. A singlet $\ket{S}$ state does not evolve under the decoupling as it is a fully symmetric state (\textbf{Fig. \ref{fig:LGextension}a}). The decoupling only alters the NV spectrum by scaling down the amplitudes of the frequency shifts as discussed in the main text. The evolution of the triplet state $\ket{T_0}$ is characterized by partial depopulation of $\ket{P_{\mathrm{R}}}$ and $\ket{P_{\mathrm{S}}}$ states and population of the double quantum $\ket{T_+}$ and $\ket{T_-}$ states (\textbf{Fig. \ref{fig:LGextension}c}). If sufficiently high decoupling strengths can be achieved, the populations of the ZQ states and the populations of the DQ states are always identical. Asymmetries in the DQ space arise due to insufficient decoupling, however if FSLG decoupling is applied instead of the more conventional continuous-wave LG decoupling, those asymmetries are alternated and thus averaged out over time (compare shaded areas in \textbf{Fig. \ref{fig:LGextension}c}). The evolution of the polarized states (\textbf{Fig. \ref{fig:LGextension}b}) is discussed extensively in the main text.

\section{Analytical Expression for DC Field Sensing of Oriented Monolayers}\label{sec:appendC}

The NV$^-$ interacts with the i$^{\mathrm{th}}$ external spin (\textit{i.e.}, a nuclear and electron spins outside of the diamond lattice) \textit{via}  a magnetic dipole-dipole interaction that can be described by a dipolar coupling Hamiltonian 
\begin{equation} 
\hat{H}_{\mathrm{dip}}^{(i)} = \frac{\mu_0 \hbar \gamma_{\mathrm{NV}}\gamma_{i}}{4 \pi}\frac{1}{r_i^3} \left( \vec{\hat{T}}\cdot \vec{\hat{S}}^{(i)} - \frac{3}{r_i^2}(\vec{\hat{T}} \cdot \vec{r}_i)(\vec{\hat{S}}^{(i)}\cdot \vec{r}_i) \right) 
\end{equation}
where $\mu_0$ is the vacuum permeability, $\gamma_{\mathrm{NV}}$ and $\gamma_{i}$ are the gyromagnetic ratios of the NV$^-$ and the external spin respectively and $\vec{r}_i$ is the  connection vector those with length $r_i$. For simplicity, we will abbreviate the constant prefactor as  
\begin{equation}
    \kappa =  \frac{\mu_0 \hbar \gamma_{\mathrm{NV}} \gamma_i}{4 \pi}
\end{equation}

Longitudinal magnetization of external spins can be probed with NV$^-$ centers in diamond  since the projection of the DC magnetic field of the partially polarized external spins onto the NV$^-$ center's z-axis induces frequency shifts $\Delta\omega_{\mathrm{A}}^{(i)}$ on the NV$^-$ center's ODMR transitions. This effect is encoded in the secular term of the dipolar coupling as elaborated on in the main text where an analytical expression for $\Delta\omega_{\mathrm{A}}^{(i)}$  is being provided. 

We now want to evaluate the maximum resulting frequency shift from multiple external spins  $\Delta\Omega_{\mathrm{A}}(V) = \sum_i \Delta\omega_{\mathrm{A}}^{(i)} = \int_V \varrho_{\mathrm{S}} \Delta \omega_{\mathrm{A}}(V)\mathrm{d}V$ by integrating over a sensing volume $V$ with a physical spin density $\varrho_{\mathrm{S}}$. We consider single NV$^-$ centers at fixed depth $d$ under a diamond surface which is covered by a homogeneous film of thickness $h$ of the spins of interest. The sensing volume can thus be approximated  as a cylinder of height $h$ and radius $R$ over which the relevant dipolar contributions have to be integrated. In a first step, the Cartesian laboratory frame coordinates have to be transformed into cylindrical coordinates following $x_{\mathrm{lab}} = R \cos{\phi_c}$, $y_{\mathrm{lab}} = R \sin{\phi_c}$ and $z_{\mathrm{lab}} = h$. For monolayers, $h = d$ and integration is only performed over $\phi_c$ and $R$
\begin{equation}
\Delta \Omega_{\mathrm{A}}(V^{\mathrm{mono}}) = \int_{0}^{R} \int_{0}^{2\pi} \varrho_{\mathrm{S}} R \Delta\omega_{\mathrm{A}}(R, \phi_c, h) \mathrm{d}\phi_c \mathrm{d}R
\end{equation}
We obtain the following expression
\begin{equation}
\Delta\Omega_{\mathrm{A}}(d , R, \theta_{\mathrm{NV}}) = \varrho_{\mathrm{S}} \kappa  \braket{\hat{S}_z^{(i)}} \frac{\pi R^2 (1 + 3\cos{2\theta_{\mathrm{NV}}})}{2(d^2 + R^2)^\frac{3}{2}}
\label{eq:analyticalmonolayer}
\end{equation}
in which $\theta_{\mathrm{NV}}$ is the angle between the surface normal and the NV$^-$ center's ZFS axis that is determined by the diamond surface termination. This expression  vanishes for non-structured surfaces ($R \rightarrow \infty$). Notably, $\varrho_{\mathrm{S}}$ is a surface density in Eq. \eqref{eq:analyticalmonolayer}, \textit{i.e.}, spins per unit area. \\

For oriented monolayers of D--B--A molecules with a stand-off distance $l$ and two oppositely polarized electron spins with a separation $s$ we can thus obtain a final expression
\begin{widetext}
\begin{equation} 
\Delta\Omega_{\mathrm{A}}(d , R, \theta_{\mathrm{NV}}, s, l) =  \pm \varrho_{\mathrm{mol}} \kappa  \frac{\pi R^2(1 + 3\cos{2\theta_{\mathrm{NV}}})}{4}\left[\frac{1}{((d+l)^2 + R^2)^\frac{3}{2}}  -  \frac{1}{((d+l+s)^2 + R^2)^\frac{3}{2}} \right]
\label{eq:dcmonodba}
\end{equation}
\end{widetext}
The sign of \eqref{eq:dcmonodba} is determined by the handedness of the donor-bridge acceptor molecule. Here, $\varrho_{\mathrm{mol}}$ refers to a molecular surface density (molecules/area) instead of a spin density. Again, the expression vanishes for unpatterned samples (\textit{i.e.}, $R \rightarrow \infty$). Analogous to the single-molecule case discussed in the main text, we define the LG-corrected frequency shift as $\Delta\Omega_{\mathrm{A}}^{\mathrm{LG}} = \cos{\theta_{\mathrm{LG}}}\Delta\Omega_{\mathrm{A}}$.

\section{Expectable Surface Densities for Example Diamond Surface Chemistries} \label{sec:appendD}

To estimate the molecular surface density $\varrho_{\mathrm{mol}}$ (in molecules/nm$^2$), we performed Monte-Carlo type simulations to model surface functionalization and molecular assembly in a way that is representative of well-defined surface chemistries for covalent anchoring of D--B--A molecules on diamond surfaces. Based on our previous work,\cite{Abendroth2022} one would first introduce chemical anchors with an average density $\varrho_{\mathrm{anchor}}$ (\textit{e.g.}, amine moieties using ammonia plasma treatment), enabling attachment of D--B--A molecules \textit{via} suitable chemical linkers in a second step. Then, $\varrho_{\mathrm{mol}}$ depends on the geometric footprints of the D--B--A molecules. In principle, anchoring will not occur on all available anchoring sites, but only such that the minimum distance between two reacting anchors $(d_{\mathrm{min}})$ is equal to the diameter of the molecule's footprint. 

In our simulations, we first generated $1 000$ random geometries of varying  $\rho_{\mathrm{anchor}}$ between $\SI{0.1}{anchors/nm^2}$ and $\SI{5}{anchors/nm^2}$ and then excluded anchoring sites until all distances are larger than $d_{\mathrm{min}}$. In analogy to the single-molecule case, we define the LG-corrected shift as $\Delta$.

\begin{figure}
\includegraphics[width=0.45\textwidth]{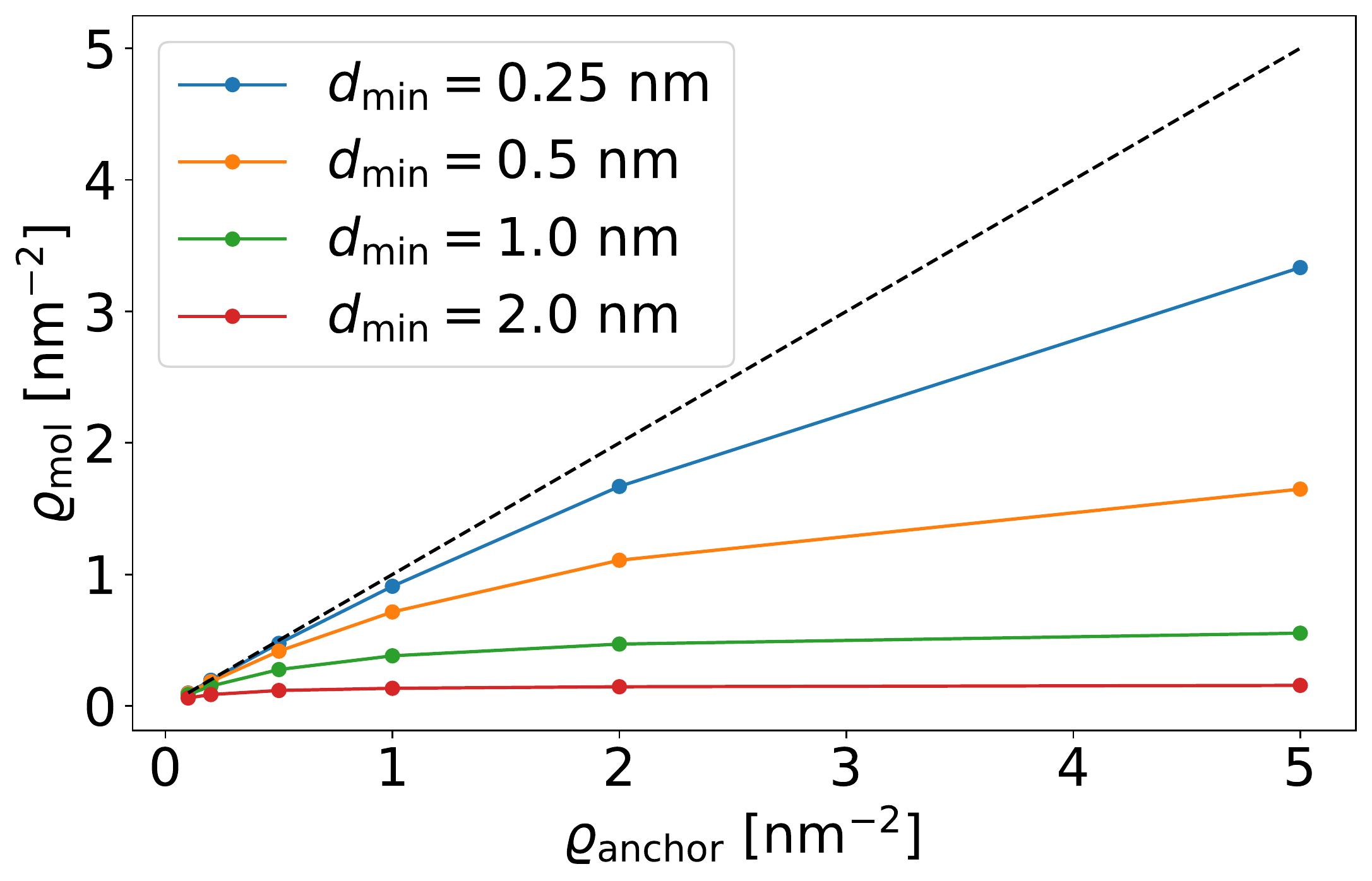}
\caption{Achievable surface densities $\rho$ of molecules with differing molecular footprints $d_{\mathrm{min}}$ ranging from $\SI{0.25}{nm}$ to $\SI{2}{nm}$ as a function of density of anchoring groups, $\rho_{\mathrm{anchor}}$, on the diamond surface. Densities were obtained in Monte-Carlo like simulations (see main text for simulation details).}
\end{figure}

\section{Details on Spin Simulations and Simulation Parameters}\label{sec:appendE}

In the following, we provide further details on the input parameters for explicit spin dynamics simulations.

\subsection{Fig. \ref{fig:LGdecoupling}}

Rotating frame simulations were performed for a single radical pair ($s=\SI{2}{nm}$, $\gamma_1=\gamma_2=\gamma_\mathrm{e}$) with an initial state $\ket{P_{\mathrm{R}}}$ ($\alpha=1$, $\beta=0$, $\Lambda=1$). The evolution under the pseudosecular dipolar coupling and FSLG decoupling, whose strength is increased from $\SI{0}{MHz}$ to $\SI{50}{MHz}$ in steps of $\SI{1}{MHz}$, was calculated for a total duration $t=\SI{1}{\upmu s}$ with time steps $\Delta t=\SI{0.05}{ns}$.  In part (b) of the figure, the time-averaged state occupations $\overline{c}_i$ and resulting time-averaged polarization $\overline{p}$ are calculated over the maximum multiple of LG decoupling periods contained within $t$. In figure (d) and (e) the first $\SI{150}{ns}$ of the time dynamics are shown for decoupling strengths of $\SI{0}{MHz}$ and $\SI{50}{MHz}$ respectively. 

\subsection{Fig. \ref{fig:singlemolecule}}

ODMR simulations were performed for a three-spin system consisting of the electron spin of an NV$^-$ center and a single radical pair ($s=\SI{2}{nm}$, $\gamma_1 = \gamma_2  =\gamma_\mathrm{e}$, $\theta_{\mathrm{RP}}=\SI{0}{\degree}$, effective depth $d=\SI{5}{nm}$) whose initial spin states are chosen as indicated in the figure legend and caption. The system Hamiltonian is composed of the  zero-field-splitting of the NV center, Zeeman interactions for all spins ($B_z=\SI{40}{mT}$), the dipolar couplings between them and $\SI{50}{MHz}$ FSLG decoupling on the electron spins of the radical pair. Frequency-switching is implemented   by reversing the sign of the frequency offset and shifting the rf phase by $\pi$.  As the Larmor frequencies of the electron spin of the NV$^-$ center and the electron spins of the radical pair differ, the simulations are being performed in the rotating frame of the NV$^-$ center and a laboratory frame for the radical pair such that the dipolar couplings from the electron spin of the NV center to the electron spins of  the radical pair  are truncated according to the high-field approximation while the full dipolar interaction is taken into account for the dipolar coupling within the radical pair. Each spectrum consists of a series of individual simulations where the NV$^-$ center is initially prepared in the $\ket{0}$ state and then subjected to a $\SI{4}{\upmu s}$ long $\pi$-pulse whose carrier frequency is swept over the resonance frequency of the NV$^-$ transition in a stepwise fashion. The observed signal ($c$) is proportional to the projection on the $\ket{0}$ state of the NV$^-$ after the $\pi$ pulse.

\subsection{Fig. \ref{fig:spinstates} (b)}
Rotating frame simulations were performed for a single radical pair ($s=\SI{2}{nm}$) with initial states $\ket{S}$ ($\alpha=0$, $\beta=0$, $\Lambda=1$), $\ket{P_{\mathrm{R}}}$ ($\alpha=1$, $\beta=0$, $\Lambda=1$) for varying g-factor differences as indicated in the legend of the figure. The time evolution under the Zeeman interaction ($B_z=\SI{40}{mT}$) and dipolar coupling was calculated for a total duration $t=\SI{150}{ns}$ with time steps $\Delta t=\SI{0.1}{ns}$.

\subsection{Fig. \ref{fig:LGextension}}

Rotating frame simulations were performed for a single radical pair ($s=\SI{2}{nm}$, $\gamma_1=\gamma_2=\gamma_\mathrm{e}$) with initial states $\ket{S}$ ($\alpha=0$, $\beta=0$, $\Lambda=1$), $\ket{T_0}$ ($\alpha=\pi/2$, $\beta=0$, $\Lambda=1$), $\ket{P_{\mathrm{R}}}$ ($\alpha=1$, $\beta=0$, $\Lambda=1$). The time evolution under the dipolar coupling and $\SI{50}{MHz}$ FSLG decoupling was calculated for a total duration $t= \SI{70} {ns}$ with time steps $\Delta t=\SI{0.1}{ns}$. Frequency-switching was implemented by reversing the sign of the frequency offset and shifting the rf phase by $\pi$.

\bibliography{aipbib}%

\end{document}